# On the Asymmetric Volatility Connectedness

Abdulnasser Hatemi-J

Department of Economics and Finance, UAE University

Email: AHatemi@uaue.ac.ae

Abstract

Connectedness measures the degree at which a time-series variable spills over volatility to other variables compared to the rate that it is receiving. The idea is based on the percentage of variance decomposition from one variable to the others, which is estimated by making use of a VAR model. Diebold and Yilmaz (2012, 2014) suggested estimating this simple and useful measure of percentage risk spillover impact. Their method is symmetric by nature, however. The current paper offers an alternative asymmetric approach for measuring the volatility spillover direction, which is based on estimating the asymmetric variance decompositions introduced by Hatemi-J (2011, 2014). This approach accounts explicitly for the asymmetric property in the estimations, which accords better with reality. An application is provided to capture the potential asymmetric volatility spillover impacts between the three largest financial markets in the world.

1. **Introduction**[1]

Investigating dynamic interaction between variables of interest is an important issue when time series variables are the focus of empirical research. Sims (1980) introduced impulse response functions and variance decompositions that are frequently used in applied research via the vector moving average version of the vector autoregressive (VAR) model. The initial approach for the identification of chocks was suggested by Sims to be Cholesky decomposition. This approach is, however, vulnerable to the way the variables enter the model. To remedy this issue Koop et. al., (1996) and Pesaran and Shin (2008) developed the generalized approach. A common deficiency of all previous approaches on measuring the effect of shock is the implicit assumption that the strength of a negative shock is the same as the strength of a positive one in absolute terms. However, this assumption does not accord with reality in many case. For seminal theoretical foundations of asymmetric behavior by financial agents see Akerlof (1970), Spense (1973) and Stiglitz (1974). The main factor behind the asymmetric structures in finance is stated to be asymmetric information. However, we argue that reason for asymmetric behavior exists even in cases in which all actors have the same information. For example, investors' response to a volatility shock might not be identical since they have heterogenous risk preferences. Some investors or financial institutions might be more risk tolerant and keep their investment positions as it is when a shock takes place in the market. While others may have lower level of risk acceptance rates and therefore change their investment positions as a cautiously rebalancing act to the occurring volatility shock. This means that there is a plausible need for measuring the impact of volatility independently conditional on if the change is positive or negative. In fact, if the asymmetric property is not considered when measuring the responses to a volatility shock, it implicitly means assuming homogeneous risk preferences for all financial actors in the market. This does not accord with reality naturally. Thus, Hatemi-J (2011, 2014) has introduced asymmetric impulse response functions and asymmetric variance decompositions for estimating the impact of a positive shock separately from the impact of a negative one. The main objective of the current paper is to extend the methodology of volatility connectedness pioneered by Diebold and Yilmaz (2012, 2014)[2] through incorporating an asymmetric structure via the asymmetric generalized variance decompositions introduced by Hatemi-J (2011, 2014). This extension provides the possibility of clearly measuring the asymmetric proportion of the forecast error variation of one variable caused by positive or negative shocks evolving from other variables. This approach is expected to accord better with reality since it accounts for the asymmetric behavior of actors in financial markets, where its existence appears to be prevailing more as a rule than an exception. There have been attempts in the existing literature for estimating asymmetric spill-over effects such as Baruník, et. al., (2016, 2017). However, this work is based on stationary variables without deterministic trend parts. However, most economic or financial variables are characterized by both a stochastic trend as well as a deterministic trend. If the variable has a unit root but it is first differenced to make it stationary, then long-run information is lost. If deterministic trend parts are neglected the data generating process of the data could be mis-specified. In the current work these issues will be dealt with to enhance the precision of the underlying inference.

---

[1] This project is funded by the CBE Annual Research Program (CARP) 2024 provided by the United Arab Emirates University, which is highly appreciated.

---

[2] For an alternative approach via the GARCH approach see Engle et. al., (1990). For a recent survey of the literature on stock market volatility see Dhingra et. al., (2024). Flores-Sosa et. al., (2022) provide a review of the literature on volatility in the foreign currency markets.

The rest of the paper is constructed as follows. The subsequent section presents asymmetric variance decompositions and thereby asymmetric measures of volatility spillover. Section 3 offers a numerical application. Concluding statements are expressed at the end.

## 2. The Asymmetric Variance Decompositions

To transform a variable that is integrated of order one with deterministic parts we follow the approach suggested by Hatemi-J (2014). Let us consider the following data generating process for a variable entitled $G_t$ that is intended for transforming:

$$G_t = c + dt + G_{t-1} + v_t \qquad (1)$$

The parametric constants $c$ and $d$ can be estimated using the least squares approach. The time trend is signified by $t$ for the sample period $t=1, \ldots, T$. The term $v_t$ is the error process that is assumed to be an identical independently distributed (IDD) random variable. Recursively substituting yields the following solution:

$$G_t = ct + \frac{t(t+1)}{2}d + G_0 + \sum_{r=1}^{t} v_r \qquad (2)$$

$G_0$ represents the initial value of the variable. The random variable $v_i$ is transformed into positive and negative elements by utilizing the definition: $v_r^+ := \max(v_r, 0)$ and $v_r^- := \min(v_r, 0)$. Via these values, the following expression is obtained:

$$G_t = ct + \frac{t(t+1)}{2}d + G_0 + \sum_{r=1}^{t} v_r^+ + \sum_{r=1}^{t} v_r^- \qquad (3)$$

Accordingly, the cumulative partial sums for $G_t$ are outlined as the following:

$$G_t^+ := \frac{ct + \left[\frac{t(t+1)}{2}\right]d + G_0}{2} + \sum_{r=1}^{t} v_r^+ \qquad (4)$$

Plus

$$G_t^- := \frac{ct + \left[\frac{t(t+1)}{2}\right]d + G_0}{2} + \sum_{r=1}^{t} v_r^- \qquad (5)$$

This transformation fulfills the following necessary condition for a correct transformation: $G_t = G_t^+ + G_t^-$. Any other variable in the VAR model can transformed in a similar way. Note that a general solution is presented here. If the integrated variable does not have any deterministic trend, then $b=0$ per definition. If there is no drift in addition to no trend, then $a=0$ and $b=0$ should be imposed. For a systematic approach of selecting the deterministic trend parts see Hacker and Hatemi-J (2010). The data can be transformed by using the statistical software component produced by Hatemi-J and Mustafa (2016) in the Visual Basic for Applications (VBA), which is compatible to the MS-Excel. This software presents a user graphical interface (GUI) that makes its use easy for all three options pertinent to the deterministic parts.

Because genuinely exogenous variables rarely exist, using the VAR model introduced by Sims (1980) is useful since it deals with all variables endogenously. This model can also be applied for measuring the dynamic interaction between variables asymmetrically according to Hatemi-J (2011, 2012). For this purpose, a separate VAR model needs to be estimated for the negative or positive components of the underlying variables. For example, consider the case in which the focus is on the relationship negative components of *m* variables. In that case, the vector $Z_t^- = (Z_{1t}^-, Z_{2t}^-, \ldots, Z_{mt}^-)$ needs to be used for estimating the following VAR(*p*) model:

$$Z_t^- = B_0^- + B_1^- Z_{t-1}^- + \cdots + B_p^- Z_{t-p}^- + u_t^- \qquad (6)$$

Here $B_0^-$ stands for an *m*×1 vector, $B_s^-$ (*s*=1, …, *p*.) is an *m*×*m* matrix, and $u_t^-$ is an *m*×1 vector of error terms. The optimal lag length, *p*, needs to be selected via the minimization of an information criterion.[3] For estimating the asymmetric variance decompositions, this VAR model needs to be presented in its moving average form as the following:

$$Z_t^- = \sum_{i=0}^{\infty} C_i + \sum_{i=0}^{\infty} K_i u_{t-i}^- \qquad (7)$$

for $t = 1, \cdots, T$. The required *m*×*m* parameter matrixes ($K_i$) can be found recursively via

$$K_i = B_1^- K_{i-1} + B_2^- K_{i-2} + \cdots + B_p^- K_{i-p}, \quad \text{for } i = 1, 2, \cdots, \qquad (8)$$

with $K_0 = I_m$ and $K_i = 0, \forall i < 0$, and $C_i = K_i B_0^-$.

---

[3] The information criterion suggested by Hatemi-J (2003, 2008), is used for this purpose. See also Mustafa and Hatemi-J (2022). Note that since the variables are integrated of the first order, an additional unrestricted lag needs to be added to the VAR model in order to account for the impact of the unit root according to Toda and Yamamoto (1995). A potential alternative is to make use of the vector error correction transformation of the VAR model if the variables are cointegrated.

The asymmetric generalized forecast error variance decomposition of a standard error shock in the $j$th equation of the VAR at time $t$ on $Z^-_{t+n}$, expressed by $AVD_{ij}(n)$, is estimated as the following:

$$AVD_{ij}(n) = \frac{\sigma_{ii}^{-1} \sum_{l=0}^{n} (e'_i K_l \Gamma e_j)^2}{\sum_{l=0}^{n} e'_i K_l \Gamma K'_l e_i} \qquad (9)$$

For $ij = 1, 2, \ldots, m$. The denotation $\Gamma$ represents the estimated variance-covariance matrix of the VAR model, that is ($\Gamma = \{\sigma_{ij}, i, j = 1, 2, \ldots, m.\}$), and $e_j$ is a $m \times 1$ choice vector with its $j$th element equal to one and zero for the rest of the elements.

A normalization is required to make the sum of the estimated variance decompositions equal to one, which can be achieved by the following:

$$\widetilde{AVD}_{ij}(n) = \frac{AVD_{ij}(n)}{\sum_{j=1}^{m} AVD_{ij}(n)} \qquad (10)$$

$\widetilde{AVD}_{ij}(n)$ represents volatility spill-over from variable $i$ on variable $j$ for horizon length $n$. likewise, the net value of the volatility spill-over is the following:

$$\widetilde{NAVD}_{ij}(n) = \widetilde{AVD}_{ij}(n) - \widetilde{AVD}_{ji}(n) \qquad (11)$$

The corresponding percentage aggregate values can be calculated as

$$\widetilde{AVD}_{i*}(n) = 100 \times \frac{\sum_{j=1}^{m} AVD_{ij}(n)}{m} \quad i \neq j \qquad (12)$$

$$\widetilde{AVD}_{*i}(n) = 100 \times \frac{\sum_{j=1}^{m} AVD_{ji}(n)}{m} \quad i \neq j \qquad (13)$$

Note that $\sum_{i,j=1}^{m} AVD_{ij}(n) = m$, which is the number of variables in the VAR model.

The directional volatility spill-over for variable $i$ for horizon $n$, denoted by $\breve{D}_{i*}(n)$, that is received from all other variables is defined as

$$\breve{D}_{i*}(n) = 100 \times \frac{\sum_{j=1}^{m} AVD_{ij}(n)}{\sum_{j=1}^{m} AVD_{ij}(n)} \quad i \neq j \qquad (14)$$

Likewise, the directional volatility spill-over from variable $i$, denoted by $\breve{D}_{i*}(n)$, that is transmitted to all other variables is expressed as

$$\breve{D}_{*i}(n) = 100 \times \frac{\sum_{j=1}^{m} AVD_{ji}(n)}{\sum_{j=1}^{m} AVD_{ji}(n)} \quad i \neq j \tag{15}$$

The design for the connectedness relationship is like the symmetric approach suggested by Diebold and Yilmaz (2012, 2014), which is presented in Table 1 for negative shocks.

**Table 1**: The Design for the Relative Spillover of Volatility (Connectedness Table).

|  | $Z_1^-$ | $Z_2^-$ | ... | $Z_m^-$ | Volatility From Others |
|---|---|---|---|---|---|
| $Z_1^-$ | $AVD_{11}(n)$ | $AVD_{12}(n)$ | ... | $AVD_{1m}(n)$ | $\sum_{j=1}^{m} AVD_{1j}(n)$ $j \neq 1$ |
| $Z_2^-$ | $AVD_{31}(n)$ | $AVD_{22}(n)$ | ... | $AVD_{2m}(n)$ | $\sum_{j=1}^{m} AVD_{2j}(n)$ $j \neq 2$ |
| ⋮ | ⋮ | ⋮ | ⋮ | ⋱ | ⋮ |
| $Z_m^-$ | $AVD_{m1}(n)$ | $AVD_{m2}(n)$ | ... | $AVD_{mm}(n)$ | $\sum_{j=1}^{m} AVD_{mj}(n)$ $j \neq m$ |
| Volatility To Others | $\sum_{i=1}^{m} AVD_{i1}(n)$ $i \neq 1$ | $\sum_{i=1}^{m} AVD_{i1}(n)$ $i \neq 2$ | ... | $\sum_{i=1}^{m} AVD_{i1}(n)$ $i \neq m$ | $\frac{1}{m}\sum_{ij=1}^{m} AVD_{ij}(n)$ $i \neq j$ |

Notes: The forecasting horizon is represented by *n*. A similar design can be created for the positive components or for other potential mixed combinations.

Note that the following equation represents the total percentage of spillover, which is an index per definition.

$$Spillover\ index = \frac{100}{m} \sum_{i,j=1}^{m} AVD_{ij}(n), \quad for\ i \neq j \tag{16}$$

This is an asymmetric version of the spillover index that was introduced by Diebold and Yilmaz (2009), which ascertains the size of spillovers of volatility shocks through all variables to the entire forecast error variance. This index can be estimated dynamically to capture the potential time-variation in its value for any sample using the different windows.

Following Diebold and Yilmaz (2012, 2014) the net value of spill-over volatility for variable $i$ to all other variables in the model for the forecast horizon $n$ is defined as the following:

$$\widetilde{ND}_i(n) = \breve{D}_{i*}(n) - \breve{D}_{*i}(n) \qquad (17)$$

A measure for net two-by-two between variables $i$ and $j$ in the VAR model can be estimated as

$$\widetilde{ND}_{ij}(n) = \left[ \frac{\widetilde{AVD}_{ij}(n)}{\sum_{r=1}^{m} \widetilde{AVD}_{ir}(n)} - \frac{\widetilde{AVD}_{ji}(n)}{\sum_{r=1}^{m} \widetilde{AVD}_{jr}(n)} \right] \times 100 \qquad (18)$$

Thus, $\widetilde{ND}_{ij}(n)$ is the distinction between gross volatility shocks conveyed from variable $i$ to variable $j$ and the reverse volatility transmission. By using this measure it is possible to find out whether any variable is net receiver of volatility from any other variable or not and by how much. These values can also be presented via network graphical illustration.

Similar estimations need to be made for the positive components. That is, equations (6)-(18) need to be estimated using the vector $Z_t^+ = (Z_{1t}^+, Z_{2t}^+, \ldots, Z_{mt}^+)$.

## 3. An Application

We apply the suggested method to capture the dynamics of the spillover effects for both positive and negative shocks for the three largest financial markets worldwide—namely the US, Euro area and the Chinese markets. Natural logarithmic values were used to account for continuously compounded impacts. The sample covers January 1999 until December 2023. The source of the data is FRED database, which is provided by the Federal Reserve Bank of St. Louis. Each variable was transformed into positive and negative components using equations (3) and (4). The VBA software component produced by Hatemi-J and Mustafa (2016) is used for transforming the data.[4] The transformed data is employed to estimate the asymmetric volatility spillover impacts through an Add-in software component in EViews created by Luvsannyam D. (2018), which is available online.

---

[4] An alternative software component for transforming the data is produced by Hatemi-J and Mustafa (2016b) in the programming language Octave.

The estimation results are produced by a VAR(2) in each case and are presented in Tables (2) and (3) using the same design that is presented in Table 1. When comparing the results, it becomes evident that the spillover impact is higher for a negative shock compared to a positive one. Thus, it is important to allow for asymmetry when measuring the volatility spillover impacts across these three financial markets. The Euro area market contributes mostly (i.e., 65.6%) to the spillover impact for rising prices. The US influences mostly (i.e., 61.4%) the spillover impact pertaining to falling prices. The Chinese market has smaller contributions to the other markets and reacts less to these markets regardless of if the prices are rising or falling. However, this impact is higher for a price decrease compared to a price increase. The Chinese market receives 0.1% of volatility from others, and it contributes by 3.9% to the volatility to other markets in case of a positive chock. The corresponding values for a negative volatility shock are 15.4% of volatility spillover to other markets and 35.7% receipt of volatility from these markets.

**Table 2**: Spillover (Connectedness) Table for Positive Shocks

|  | China | Euro | US | From Others |
|---|---|---|---|---|
| China | 99.9 | 0 | 0.1 | 0.1 |
| Euro | 0.6 | 97.9 | 1.4 | 2.1 |
| US | 3.3 | 65.6 | 31.2 | 68.8 |
| Contribution to others | 3.9 | 65.6 | 1.5 | 71 |
| Contribution including own | 103.8 | 163.5 | 32.7 | 23.70% |

**Table 3**: Spillover (Connectedness) Table for Negative Shocks

|  | China | Euro | US | From Others |
|---|---|---|---|---|
| China | 64.3 | 18.5 | 17.3 | 35.7 |
| Euro | 5.3 | 50.6 | 44.1 | 49.4 |
| US | 10 | 38.4 | 51.6 | 48.4 |
| Contribution to others | 15.4 | 56.9 | 61.4 | 133.6 |
| Contribution including own | 79.6 | 107.4 | 112.9 | 44.50% |

The volatility spillover index value for the three markets is 44.5% for the negative shock while it is 23.7% for the positive shock using equation (16). These volatility spillover index values are also calculated for 200-month windows when $n=10$, which are presented in Figures (1) and (2) in order to capture time variation. These figures show once again that the asymmetric impact is important to consider since the time path of this volatility spillover index is significantly different for a positive shock compared to a negative one. For positive shocks, the index value is relatively stable around 30% except for the pandemic period for 2019-2020 years that is slightly higher. However, the index value has a clear upward time trend starting at around 37% and reaching around 48% for a negative shock. These asymmetric volatility linkage results have important implications for investors and financial instructions with different positions in the underlying investment depending on whether a long or a short position is undertaken. The results could also be used by policy makers to design appropriate strategies for damping potential external shocks especially when the markets are falling to avoid or reduce the resulting contagion effects.

Figure 1: Spillover (Connectedness) Index for the Positive Shock (200 month windows, 10 step horizons).

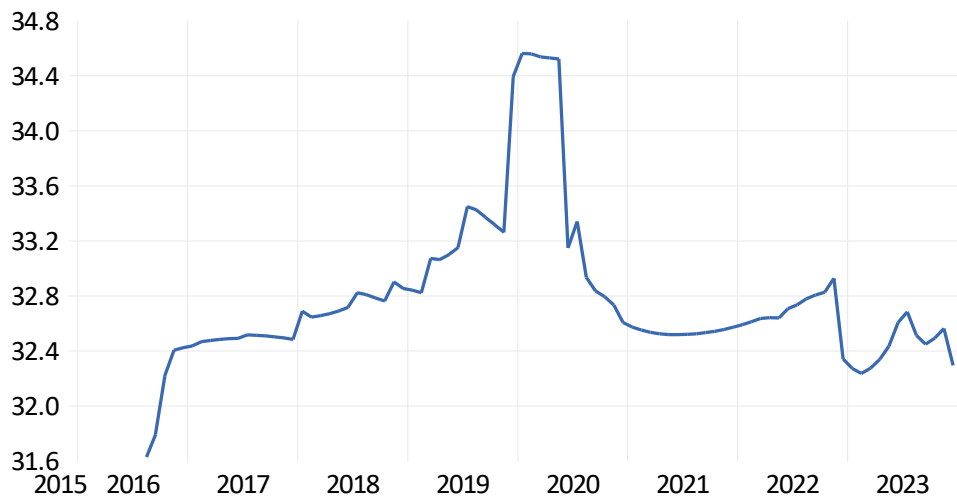

Figure 2: Spillover (Connectedness) Index for the Negative Shock (200 month windows, 10 step horizons).

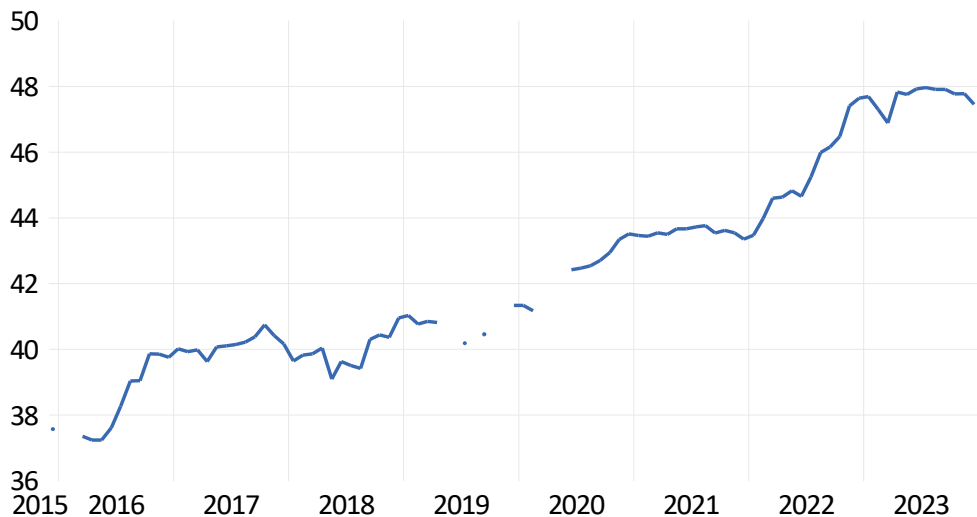

The symmetric volatility spillover impacts are also estimated and presented in Table 4 and Figure 3. These results show that the asymmetric impacts indeed prevail. For example, the spillover index value for the symmetric case is 38.60%, which is a different amount compared to the values for the asymmetric cases since this value is 23.70% for a positive shock and it is 44.50% for a negative shock.

**Table 4**: Symmetric Spillover (Connectedness) Table.

|  | China | Euro | US | From Others |
|---|---|---|---|---|
| China | 75.5 | 13.2 | 11.3 | 24.5 |
| Euro | 3.3 | 54.3 | 42.3 | 45.7 |
| US | 6.8 | 38.9 | 54.3 | 45.7 |
| Contribution to others | 10.2 | 52 | 53.6 | 115.8 |
| Contribution including own | 85.7 | 106.4 | 107.9 | 38.60% |

Figure 3: Symmetric Spillover (Connectedness) Index (200 month windows, 10 step horizons).

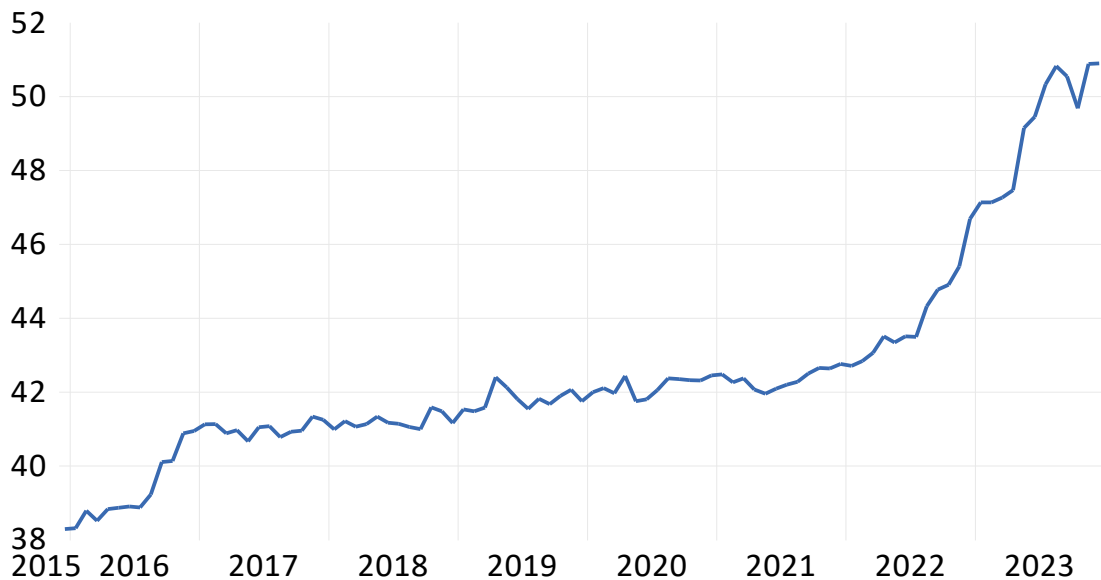

## 4. Concluding Remarks

Volatility as a measure of financial risk is of fundamental importance for investors, financial institutions, and policy makers. Measuring the potential volatility spillover impacts is increasingly gaining attention and practical necessity in line with the increasing internationalization and consequently the enhanced connection between markets across the borders. The methodological contributions of Diebold and Yilmaz (2009, 2012, 2014) are useful for this purpose, which is also demonstrated by the huge number of applications in empirical research.

The Diebold and Yilmaz approach is symmetric per construction, however, since it does not account for the potential distinctive impact of a negative shock compared to a positive one. It is extensively agreed in the literature that asymmetric impacts prevail, especially in the financial markets. There are several reasons for this asymmetric structure. The most prominent reason for this issue is claimed to be asymmetric information. However, the asymmetric reaction to a change in the market can prevail even if the information is symmetric since investors have different preferences towards risk. Thus, it is reasonable to allow for potential asymmetric impacts when measures for volatility spillover impacts are estimated, which is the main objective of this paper. Hence, Diebold and Yilmaz (2009, 2012, 2014) methodology is extended by making it asymmetric. This paper demonstrates how a variable with potential stochastic and deterministic trend parts can be transformed into positive and negative components, which can be used for estimating the asymmetric volatility spillover impacts. The suggested approach makes it operational to estimate the volatility spillover impact with regard to both a negative shock and a positive one. The estimation results based on this approach is expected to accord better with reality. This distinction between a positive and negative chocks also has important practical implications. For example, the source of risk for an investor with a long position in the asset is decreasing prices and hence this investor is concerned about negative shocks. While the source of risk for an investor with a short position in the asset is increasing prices and thereby the concern is focused on positive shocks for such an investor. Likewise, the policy makers are mainly concerned about the negative shocks, especially during a financial crisis in order to hedge against the potential contagion effect as much as possible.

An application is provided for estimating symmetric as well asymmetric spillover impacts between the three largest financial markets in the world—namely the US, the Euro area and China. The empirical results show that volatility spillover impacts between these markets is indeed asymmetric. The volatility spillover percentage is much higher for a negative shock compared to a positive one. The time variation for these values also indicates that the variation in the spillover index is higher for a negative shock versus a positive one across time. These results could be informative to investors, financial institutions and policy makers and thereby impact their decision making for the better.